\documentclass[12pt]{iopart}

\usepackage{iopams}
\usepackage{epsfig}
\begin{document}
%%%%%%%%%%%%%%%%%%%%%%%%%%%%%%%%%%%%%%%%%%%%%%%%%%%%%%%%%%%%%%%%%%%
\def\be {\begin{equation}}
\def\ee {\end{equation}}
\def\nn {\nonumber}
\def\bea {\begin{eqnarray}}
\def\eea {\end{eqnarray}}
\def\k {|{\vec k}|}
\def\q {|{\vec q}|}
\def\ks {{|{\vec k}|}^\ast}
\def\qs {|{{\vec q}|}^\ast}
\def\mqs {M_q^\ast}
\def\eqs {E_q^\ast}
\def\eq {E_q}
\def\mks {M_k^\ast}
\def\ek {E_k}
\def\eks {E_k^\ast}
\def\mp {m_\pi}
\newcommand{\bef}{\begin{figure}}
\newcommand{\eef}{\end{figure}}
\newcommand{\ra}{\rightarrow}
\newcommand{\gm}{\gamma^\mu}
\newcommand{\gf}{\gamma^5}
\newcommand{\N}{\bar{N}}
\newcommand{\del}{\partial}
\def\bt{{\bf{\tau}}}
\def\ba{{\bf{a_1}}}
\def\br{{\bf{\rho}}}
\def\bp{{\bf{\pi}}}
\def\cl{{\cal{L}}}
%%%%%%%%%%%%%%%%%%%%%%%%%%%%%%%%%%%%%%%%%%%%%%%%%%%5
\title[]{Thermal Radiation from Heavy Ion Collisions at RHIC}

\author{Jan-e Alam}

\address{Variable Energy Cyclotron Centre, 1/AF Bidhan Nagar,
Kolkata 700 064, INDIA} 

%\ead{jane@veccal.ernet.in}

\begin{abstract}
The  direct photon spectrum measured by the
PHENIX collaboration in $Au + Au$ collisions at
$\sqrt{s_{NN}}=200$ GeV has been analyzed. It has been shown that
the data can be reproduced reasonably well by assuming
a deconfined state of thermalized quarks and gluons.
The effects of the equation of state on the value of the initial
temperature have been studied.
The modifications of hadronic properties at non-zero
temperature have been taken in to account.
\end{abstract}

\section{Introduction}
Collisions between two nuclei at ultra-relativistic energies
produce charged particles - either in the hadronic or in the
partonic state, depending on the collision energy.
Interaction of these charged particles produce  real and
virtual photons. Because of their nature of interaction, the
mean free path of photons  
is very large compared to the size of the system
formed after the collision. Therefore, photons
can be used as an efficient tools to understand the 
initial conditions of the state where it is created~\cite{emprobes}.  
The purpose of the present work
is to analyze this experimental data obtained by 
PHENIX collaboration~\cite{phenix} 
for $Au +Au$ collisions at $\sqrt{s_{NN}}=200$ GeV 
and infer the initial
temperature ($T_i$) of the system formed after the collisions. 

\section{Photons from pQCD Processes}
The hard collisions of initial state partons in the colliding nuclei 
produce photons with high transverse momentum ($p_T$).
This contributions can be estimated by  
pQCD.  We use the  next to leading order (NLO) predictions  
of Ref.~\cite{gordon} for $pp$ collisions
and scale it up by the number of binary collisions for $Au+Au$ interactions 
to obtain
the pQCD contributions to the direct photons.
\section{Thermal Photons}
We assume here that quark gluon plasma (QGP) is formed initially which 
then expands, cools,
and reverts back to hadronic matter and finally freezes out  
at a temperature ($T_f$). Therefore, evaluation of photon
spectra both from QGP and hadronic matter is required.

\subsection{Thermal Photons from QGP}
The photon emission rate from QGP due to Compton ($q(\bar{q})g\rightarrow 
q(\bar{q})\gamma$) and annihilation ($q\bar{q}\rightarrow g\gamma$) processes 
is  evaluated in~\cite{kapusta,baier} by using Hard Thermal Loop (HTL) 
approximation.  Later it was shown~\cite{auranche1}
that photon production  from the reactions,
$gq\rightarrow gq\gamma$, $qq\rightarrow qq\gamma$,
$qq\bar{q}\rightarrow q\gamma$ 
and $gq\bar{q}\rightarrow g\gamma$ contribute in the same order
as annihilation and Compton processes.
The complete calculation of photon
emission rate from QGP to order $\alpha_s$ 
is completed
by resuming ladder diagrams in the effective theory~\cite{arnold}. 
This result has been used in the present work.

\subsection{Thermal Photons from Hadrons}
For the photon production from hadronic matter all the possible
reactions and decays involving 
$\pi$, $\rho$, $\omega$, $\eta$ and $a_1$ have been considered 
(see~\cite{we1,we2,we3} for details).
Photons from the decays of $\pi^0$, $\eta$ etc are subtracted from the data 
and hence is not discussed here. 

Various experimental~\cite{metag,djalali} and theoretical 
results~\cite{rappqm06,rw,annals,brphysrep269} suggest that 
the spectral functions of hadrons are modified in a hot and dense 
nuclear environment. 
There is no consensus on the nature of the vector meson modification
in matter - pole shift or broadening - both experimentally
and theoretically. 
In the present work we use Brown-Rho (BR) scaling scenario~\cite{brown}
for in-medium modifications of hadronic masses
to indicate how far the value of initial temperature is
affected when the reduction of the hadronic mass is incorporated
in evaluating the photon spectra. 

Transverse momentum distribution of photons produced from the 
fragmentation of fast quarks propagating 
through QGP remains almost unaffected due its energy loss~\cite{zakharov}.
Photons produced  due to interactions between thermal  and 
non-thermal partons~\cite{fries} 
have been neglected in the current analysis.

\section{Space Time Evolution}
(2+1) dimensional~\cite{hvg} (ideal) relativistic hydrodynamics
with longitudinal boost 
invariance~\cite{bjorken} and cylindrical symmetry has been used 
for the space time evolution. 
We take initial thermalization
time  $\tau_i=0.2$ fm/c, $T_i=400$ MeV and the number of flavours, 
$N_F = 2.5$.  These values reproduce
the measured total  multiplicity, $dN/dy\sim 1100$.  
The initial energy density and  radial velocity profiles 
are taken as:
$\epsilon(\tau_i,r)=\epsilon_0/[1+e^{(r-R_A)/\delta}]$\,
and\,
$v(\tau_i,r) = v_0\left[1-1/[1+e^{(r-R_A)/\delta}]\right]$.Here
$\delta$ is the surface thickness and $R_A$ is the radius of the
colliding nuclei.

Two types of equation of state (EOS) are used 
to study the photon spectra:
(I)  Bag model EOS has been used for QGP.
For the hadronic matter all the
resonances with mass
$<$  2.5 GeV $/c^2$ has been considered.
(II) Results from lattice QCD~\cite{karsch02}
has also been used 
to show the sensitivity of the results on the EOS.

\section{Results}
\begin{figure*}[htb]
\epsfig{file=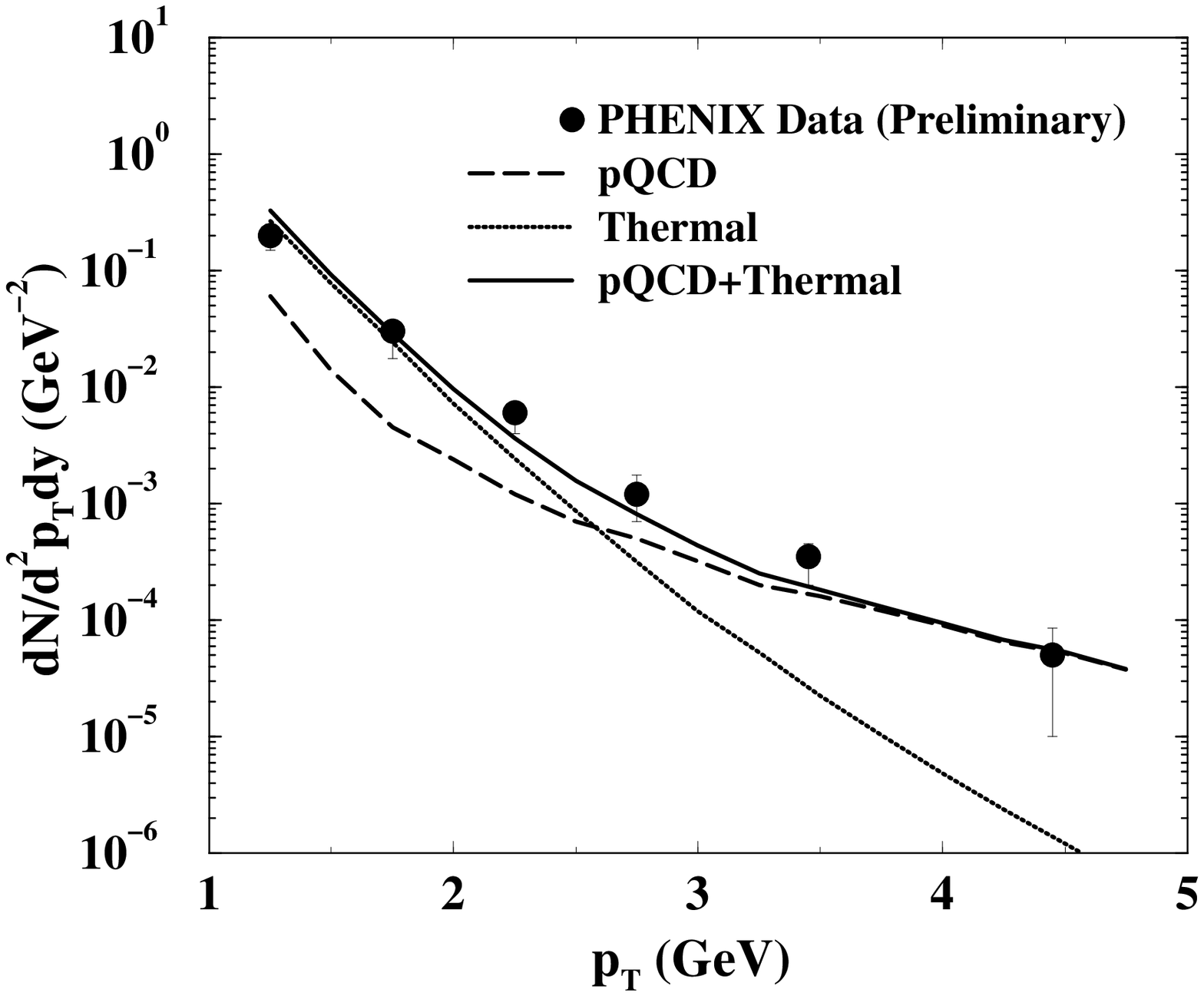, width=7.5cm}\epsfig{file=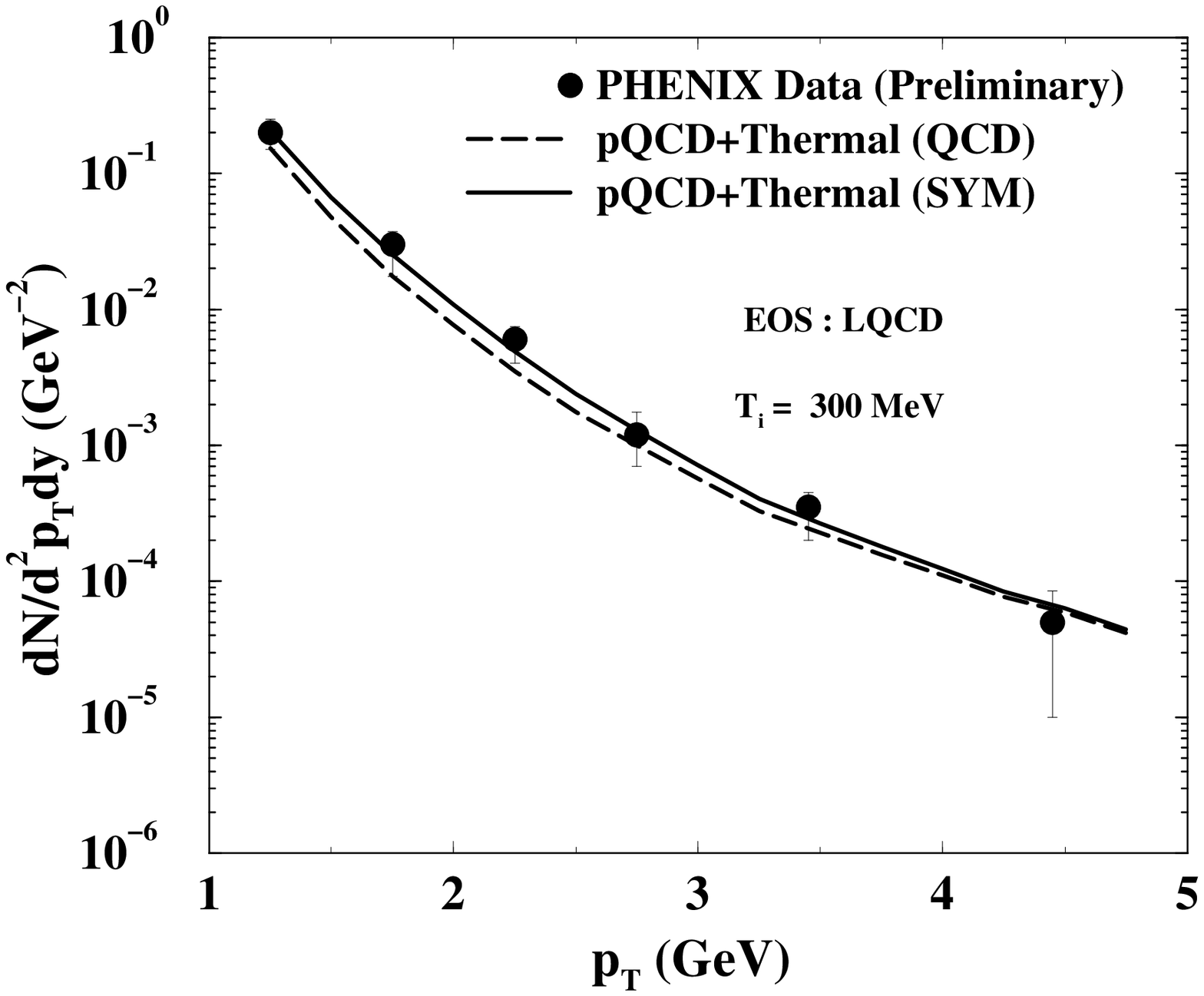, width=7.5cm}
\vspace*{-0.5cm}
\caption{\label{fig1}
Left panel:  Solid  line depicts the total [pQCD (dashed) + thermal (dotted)] 
photon yield. The value of $T_i$ = 400 MeV, 
$\tau_i=0.2$ fm/c and  EOS (I) is used here.  
Photon production rate from QGP is taken from~{\protect\cite{arnold}}.
Right panel:  The total photon yield when EOS
from lattice QCD is used.
Photon production rate from QGP is taken from~{\protect\cite{arnold}}
(dashed) and from SYM~{\protect\cite{sym}} (solid line).
}
\end{figure*}

Photon spectra is evaluated with the initial conditions
mentioned above. The value of the
freeze-out temperature $T_f=120$ MeV~\cite{jjr} has been 
fixed from the $p_T$ spectra of pions and kaons~\cite{pikphenix}.
The value of the transition temperature, $T_c=190$ MeV~\cite{cheng}
is taken here.
The resulting photon spectra is contrasted 
with the recent PHENIX measurements of direct
photons in Fig.~\ref{fig1}. The data is
reproduced with $T_i=400$ MeV and $\tau_i=0.2$ fm/c. 
The value of $T_i$ is smaller compared to the value obtained 
in Ref.~\cite{peress}.  Because the reduction of hadronic masses in
a thermal bath increases their abundances and hence the rate of 
photon emission gets enhanced. 
Therefore, to pinpoint  the actual initial temperature through photon 
spectra it is imperative to understand the properties
of hadrons in hot and dense environment.
This  may be compared with the value of $T_i\sim 200$ MeV
obtained from the analysis~\cite{ja1} of WA98 data~\cite{wa98} at SPS energy.

The photon spectra is also evaluated using EOS from lattice QCD.
It is seen (right panel of Fig.~\ref{fig1}) 
that the data can be reproduced 
with lower
$T_i\sim 300$ MeV (and hence larger 
$\tau_i\sim 0.5$ fm/c). 
This is so because for type II EOS  the space time evolution 
of the hadronic phase is
slower than type (I).
It may be mentioned at this point that the photon emission
rates obtained in~\cite{arnold}  are valid in the weak coupling
limit. However, the QGP formed after $Au + Au$ collisions
at RHIC energy could be strongly coupled and the  photon
production from strongly coupled QGP is not available from thermal
QCD.  Therefore, result from ${\cal{N}}=4$ Supersymmetric Yang Mills (SYM) 
theory~\cite{sym}
has been  used to estimate the photon production 
from QGP phase in the strong coupling limit. 
The rate obtained in this case could be treated as a
upper limit of photon production from QGP.  The results obtained
in this case are compared with that from thermal  QCD in 
right panel of Fig.~\ref{fig1}. 
Photons from SYM is enhanced by about $20\%$ as compared to thermal
QCD in the $p_T$ region $\sim 2 $ GeV.

%%%%%%%%%%%%%%%%%%%%Fig.2%%%%%%%%%%%%%%%%%%%%%%%%%%%%%%%%%%%%%
%%%%%%%%%%%%%%%%%%%%%%%%%%%%%%%%%%%%%%%%%%%%%%%%%%%%%%%%%
\section{Summary}
In summary, we have analyzed the direct photon data measured
by PHENIX collaboration for $Au + Au$ collisions at $\sqrt{s_{NN}}=200$ GeV.
The data can be reproduced by assuming a deconfined state of quarks
and gluons with $T_i\sim 400$ MeV.
However, for EOS from lattice QCD the data can be explained for
lower value of $T_i\sim 300$ MeV. Photon productions from thermal QCD
and ${\cal{N}}=4$ SYM have  been used for the analysis of data.
In both the cases similar values of the initial temperatures are
obtained.

{\bf Acknowledgment:} The author would like to thank Jajati K. Nayak, 
Pradip Roy, Abhee K. Dutt-Mazumder and Bikash Sinha for fruitful collaboration.

\end{document}